\documentclass{article}

\usepackage{PRIMEarxiv}

\usepackage[utf8]{inputenc} 
\usepackage[T1]{fontenc}    
\usepackage{hyperref}       
\usepackage{url}            
\usepackage{booktabs}       
\usepackage{amsfonts}       
\usepackage{nicefrac}       
\usepackage{amsmath}
\usepackage{microtype}      
\usepackage{lipsum}
\usepackage{fancyhdr}       
\usepackage{graphicx}       
\usepackage{amssymb}
\usepackage{amsmath}
\usepackage{subcaption}
\usepackage{graphicx}
\usepackage{booktabs}
\usepackage{multirow}
\usepackage{float}
\usepackage{wrapfig}
\usepackage{array}
\usepackage{comment}
\graphicspath{{media/}}     

\title{DGD-cGAN: A Dual Generator for Image Dewatering and Restoration
}

\author{
  Salma Gonzalez-Sabbagh\\
  School of IT \\
  Deakin University\\
  Victoria, Australia\\
  \texttt{sgonzalezsabbag@deakin.edu.au} \\
   \And
  Antonio Robles-Kelly\\
  Defence Science and Technology Group\\
  Edinburgh, South Australia\\
  \texttt{antonio.robles-kelly@deakin.edu.au} \\
  \AND
  Shang Gao\\
  School of IT \\
  Deakin University\\
  Victoria, Australia\\
  \texttt{shang.gao@deakin.edu.au} \\
}

\begin{document}
\maketitle

\begin{abstract}
Underwater images are usually covered with a blue-greenish colour cast, making them distorted, blurry or low in contrast. This phenomenon occurs due to the light attenuation given by the scattering and absorption in the water column. In this paper, we present an image enhancement approach for dewatering which employs a conditional generative adversarial network (cGAN) with two generators. Our Dual Generator Dewatering cGAN (DGD-cGAN) removes the haze and colour cast induced by the water column and restores the true colours of underwater scenes whereby the effects of various attenuation and scattering phenomena that occur in underwater images are tackled by the two generators. The first generator takes at input the underwater image and predicts the dewatered scene, while the second generator learns the underwater image formation process by implementing a custom loss function based upon the transmission and the veiling light components of the image formation model. Our experiments show that DGD-cGAN consistently delivers a margin of improvement as compared with the state-of-the-art methods on several widely available datasets.
\end{abstract}


\section{Introduction}\label{introduction}
Underwater computer vision has attracted increasing attention in the research community due to recent advances in underwater platforms such as rovers, gliders and autonomous underwater vehicles (AUVs). These advances now make the acquisition of vast amounts of imagery and video possible for applications such as biodiversity assessment and protection \cite{Boudhane:2016}, aquaculture \cite{Mohamed:2020}, underwater human-made object detection \cite{Rizzini:2015}, infrastructure inspection \cite{Martin:2020} and rescue and surveillance \cite{Martija:2020,Karlekar:2010}. 

Note underwater environments are somewhat unique due to the optical properties and radiative transfer that result in noticeable colour casts, image degradation, loss of contrast and low-light conditions often found in underwater scenes. This is due to the optical effects inherent to light propagation in water columns, whereby light is exponentially attenuated and scattered depending upon the concentrations of particulate and dissolved matter, the distance and angle of view \cite{Duntley:1957,Duntley:1963,Carlevaris:2010,Akkaynak:2018}. Therefore, most underwater images are covered by a blue or greenish colour tinge, appearing blurred or distorted. As a result, colour restoration and dehazing of underwater images have attracted considerable attention in the computer vision  \cite{Akkaynak:2019} and machine learning communities \cite{Skinner:2017}.

Despite effective, existing approaches to colour correcting underwater images often resort to the depth information to estimate the absorption and scattering in the water column  \cite{Akkaynak:2019}. They employ underwater and in-air images to make use of these as the reference required for the colour correction training \cite{Cho:2020} or disregard the image formation model altogether by adopting an end-to-end approach \cite{Wang:2022,Islam:2020}. In this paper, we present a two-generator approach based upon conditional Generative Adversarial Networks (cGANs) \cite{Mirza:2014} for underwater colour correction. Departing from the underwater image formation model (UIFM) in \cite{Duntley:1963,Duntley:1957}, we show how the two generators can be used to enhance underwater scenes. The first generator takes the underwater image as input and delivers the predicted radiance as output. The second generator learns how to estimate the underwater transmission and veiling light by making use of a loss function based upon the UIFM as a consitency constraint. As a result, our two-generator architecture is able to cope with wide variety of attenuation and scattering phenomena occurring in underwater images. Experiments show that our network can be trained on a small dataset of real-world images. Our contributions are summarized as follows:
\begin{itemize}
	\item We propose a DGD-cGAN comprising two generators and one discriminator. This network can perform underwater image correction devoid of input depth information or water column parameters. The dual generators are designed to cope with wide variety of attenuation and scattering phenomena occurring in underwater images.
	\item We jointly optimise two objectives: the consistency of dewatered image with their underwater scene as constrained by the UIFM in \cite{Duntley:1963} and the colour restoration. By optimising them simultaneously, we can train our model on a small set of real-world images with a good correction quality.
	\item We extensively evaluate our DGD-cGAN on five datasets and compare our method with state-of-the-art alternatives spanning from histogram equalisation (HE) to GAN-based methods. The results show that our DGD-cGAN consistently delivers a margin of improvement over the alternatives on all datasets.
\end{itemize}

\section{Related work} \label{related_work}

As mentioned earlier, obtaining clear natural underwater images is challenging due to the degradation, dimensional distortions and blurring effects \cite{Pei:2018} caused by absorption and scattering across the water column. As a result, over the last two decades, there has been increasing interest in underwater image dehazing \cite{Carlevaris:2010}, restoration \cite{Islam:2020}, colour correction \cite{Ancuti:2017:locally} and depth recovery \cite{Asano:2020}. This has been further motivated by the importance of photometric invariance for underwater computer vision tasks.

Therefore, it is not surprising that several approaches have been proposed for improving the quality of underwater images, where the main aim is recovering colour corrected imagery \cite{Henke:2013,Bianco:2015}. This is often effected by using reflectance modelling and radiative transfer theory as related to the UIFM. For instance, Akkaynak and Treibitz \cite{Akkaynak:2018} introduced a revised UIFM that takes into account the angle of sight and the distance between the object and the sensor for calculating the attenuation coefficients. Building on this model, they later on developed the Sea-thru approach in \cite{Akkaynak:2019} for recovering the in-air image of underwater scenes. In \cite{Akkaynak:2019}, the backscattering is estimated using the Dark Channel Prior (DCP) \cite{He:2010DCP} with depth information from RGBD (red, green, blue and depth) data, whereas the diffuse attenuation is computed using the local space average colour \cite{Ebner:2009}. In a related development, Carlevaris-Bianco et al. \cite{Carlevaris:2010} presented a method for underwater image dehazing based on the UIFM developed by Duntley \cite{Duntley:1963}. Based upon the notion that the light absorption in underwater scenes is higher in the red channel than in the green and blue channels, they compared the channel maximum intensities to estimate the depth using the DCP. In this manner, the colour cast can be removed by estimating the maximum posterior probability of the scene reflectance using Markov Random Fields \cite{Marroquin:1987}. 

Other approaches are phenomenological in nature. For instance, in \cite{Hitam:2013} the authors employed Contrast Limited Adaptive Histogram Equalisation (CLAHE) for underwater image enhancement. In their approach, the RGB (red, green and blue) and HSV (hue, saturation and value) colour spaces are combined via normalisation of the RGB outputs and transformation of the HSV to RGB so as to reduce brightness and noise in the images using the Euclidean norm. Ancuti et al.\cite{Ancuti:2017} proposed a fusion based method for enhancing underwater images using a white balance algorithm. They compensated the loss of the red intensity channel by adding the green channel-to-blue ratio to the lowest red channel image values. It aims at using the blue intensity channel to overcome the green cast often found in underwater imagery. In their method, the Gray World (GW) \cite{Finlayson:1998,Buenaposada:2001} algorithm is used to estimate the global illumination.

Note that the  methods above do not employ neural architectures rather employ physics-based vision or image processing techniques to recover the colour corrected underwater image. Deep learning methods have also been applied in underwater environments, whereby Convolutional Neural Networks (CNNs) and GANs are the most common architectural choices for colour correction and enhancement. Li et al.\cite{Skinner:2017} built WaterGan to generate underwater images from in-air images and their depth map using a generative adversarial network. They incorporate the attenuation, backscattering and camera characteristics  to train the generator and employ synthetic underwater scenes for training. WaterGan is based upon two CNNs reminiscent of SegNet \cite{Badrinarayanan:2017}. A similar approach was presented in \cite{Wang:2019}, where the generator used is a U-Net \cite{Ronneberger:2015} model trained using backscattering, forward scattering and attenuation estimates. Fabbri et al. \cite{Fabbri:2018} proposed an Underwater GAN (UGAN) as a means to address the visibility issues often encountered by AUVs. UGAN is also based on \cite{Isola:2017} and uses a U-Net generator and a PatchGAN discriminator for image enhancement. The authors in \cite{Wang:2019} used Cycle-consistent GANs (CycleGANs) \cite{Zhu:2017} to generate underwater image pairs with and without distortion. Li et al. \cite{Li:2019:WaterNet} proposed Water-Net, a fusion-based U-Net and residual network for underwater image enhancement. Water-Net is a CNN that employs white balance (WB) \cite{Ancuti:2012} for colour cast removal, HE \cite{Reza:2004} for contrast improvement and gamma correction (GC) for low-light conditions.   


It is worth noting that the methods above did not include the UIFM into the optimisation objective but rather employed colour constancy or directly incorporated the attenuation, backscattering, depth or camera characteristics to train the generator. None of the above methods employed multiple generators either. There are, however, methods that employed two generators. Hu et al. \cite{Hu:2021} enhanced underwater images by designing a dual-generator deep network based on the HSV colour space.  Maniyath et al. \cite{Maniyath:2021l} also introduced a network with two generators for underwater image dehazing that translated in-air images to an underwater domain for later removal of the colour cast. Note, however, that these methods are quite different to ours. In \cite{Hu:2021}, one of their generators removes the colour cast and the other improves the contrast, whereby their output is concatenated and converted to RGB before being fed to the discriminator. The method in \cite{Maniyath:2021l} is such that the two discriminators operate on restored and degraded images. In \cite{Park:2019}, the authors also proposed a GAN with two generators. In contrast to DGD-cGAN, their method uses two discriminators for dehazing, colour correction and feature extraction. Similarly, the restoration GAN presented in \cite{Du:2021} is based upon two discriminators and two generators, where one generator enhances the underwater image and the other degrades it before being fed to the two discriminators. Furthermore, all these GAN architectures with more than one generator are based on CycleGANs. That is another major difference to our DGD-cGAN, which employs a conditional GAN (cGAN) architecture.

\section{Underwater Light Propagation}
Light propagation in underwater environments is affected by the index of refraction, water molecules, particulate and dissolved matter, and marine flora and fauna. All these elements attenuate the light both due to its absorption and scattering. Recall absorption occurs when the light penetrates the water and is converted into another type of energy. Longer wavelengths, which carry the redish colours, are typically absorbed as soon as the water travels 1m in depth, whereas shorter wavelengths go deeper. Scattering occurs when a beam of light hits an element and is deviated from its straight trajectory, going backward or forward \cite{Pope:1997,Mobley:1994,Dekker:2002,Mishchenko:2002,Wozniak:2007}. As a result, underwater images are covered with a blue or greenish tinge and are distorted, blurred or exhibit low contrast. 

Thus, in underwater imaging, the modelling of image formation process has played an important role as a means to the computation of a colour corrected, photometrically invariant imagery to be obtained and used for computer vision tasks such as recognition, segmentation and shape recovery. One of the most often used models is that proposed by Duntley \cite{Duntley:1963,Duntley:1957} and later further explored by Akkaynak and Treibitz \cite{Akkaynak:2018}. This is a physics-based model extensively used in underwater computer vision, where depth $z$, distance $r$ between the camera centre and the object under consideration, and the viewer's direction defined by the zenith $\theta$ and azimuth $\phi$ angles are used to compute the light attenuation. With these ingredients, in Duntley's model \cite{Duntley:1963,Duntley:1957} the image irradiance is given by
\begin{equation}\label{eq3}
\begin{split}
    {_t}{N} {_r}(z,\theta,\phi)= {_t}{N} {_0}(z,\theta,\phi)\exp[-\alpha(&z)r]
    + N(z_{t},\theta,\phi)\exp[+K(z,\theta,\phi)r \cos{(\theta)}]\\&\times\{1-\exp[-\alpha(z)r+K(z,\theta,\phi)r \cos(\theta)]\},
\end{split}
\end{equation}   
where ${_t}{N}{_r}(z,\theta,\phi)$ is the underwater image, ${_t}{N}{_0}(z,\theta,\phi)$ is the radiance from the object, $N(z_{t},\theta,\phi)\exp[+K(z,\theta,\phi)r \cos(\theta)]$ is the veiling light and $K(z,\theta,\phi)$ is the diffuse attenuation related to the angle of view. 

The expression in Equation \ref{eq3} is often rewritten in a simplified form and used in underwater computer vision research as follows \cite{Narasimhan:2002}
\begin{equation}\label{eq1}
\mathbf{I} = \mathbf{J}\odot\mathbf{T} + \mathbf{A}\left(1-\mathbf{T}\right),
\end{equation}
where $\odot$ denotes the Hadamard, i.e. elementwise, product, $\mathbf{I}$ is the underwater image, $\mathbf{J}$ is the reflected radiance from the object, $\mathbf{A}$ is the scattered light, known as veiling light or airlight, and $\mathbf{T}$ is the transmission of light. Simplified in this manner, the attenuation along the water column is exponential in nature, given by 
\begin{equation}\label{eq2}
T(u) = \exp{(-\beta d(u))},
\end{equation}
where $u$ is the pixel under consideration, $\beta$ is the attenuation coefficient and $d(u)$ is the distance between the camera and the object at pixel $u$.

The main difference between both equations becomes evident by setting the attenuation in Equation \ref{eq1} to $\exp[-\alpha(z)r]$ and setting $K(z,\theta,\phi)$ to zero in Equation \ref{eq3}. Note that in water, the scattering is a function of depth, while the absorption is depth and wavelength dependant. Moreover, when the angle of view is horizontal with respect to the object, $K(z,\theta,\phi)=0$ \cite{Duntley:1963}.

\section{DGD-cGAN}
Underwater imaging presents a number of challenges related to the environment in which it operates. For instance, depth acquisition together with scattering and absorption measurement is a difficult task and often requires the use of special devices \cite{Jaffe:2001,Yamashita:2007}. Because of this, datasets in underwater computer vision often lack ground truth, depth information and other kinds of metadata \cite{Yuan:2022,Zhao:2021}. As a result, we aim at estimating the transmission and veiling light, including the diffuse attenuation coefficient instead of assuming a horizontal pose and setting it to zero. We do this by taking advantage of the known ground truth and making use of a cGAN that trains two generators and one discriminator. 

Here, we use a CNN model based on U-Net \cite{Ronneberger:2015} for both generators $G_1$ and $G_2$ and PatchGAN's \cite{Isola:2017,Li:2016:MarkovianDisc} discriminator. Note that U-Net has been used in recent underwater research \cite{Fabbri:2018,Islam:2020} mainly due to the fact that the input and output have similar structure and high resolution. U-Net shape downsamples the input and upsamples output features, hence preserving the image resolution by employing an encoder-decoder architecture with mirrored layers where the encoder outputs are concatenated with those of the decoder. 

Despite our network having some similarities to the image-to-image translation approach in \cite{Isola:2017}, i.e. PatchGAN, and being based upon U-Net \cite{Ronneberger:2015} for its generators, there are a number of notable differences. As related to the U-Net architecture  presented in \cite{Ronneberger:2015}, there is twofold. Firstly, we include the skip connections as in \cite{Isola:2017,Chen:2018:CVPR}; secondly, we add deep residual units between them as in \cite{He:2016,Zhang:2018:Resunet}. Thus, our DGD-cGAN modularity is BatchNorm-ReLU-Convolution-ResU. By contrast, the PatchGAN in \cite{Isola:2017} is Convolution-BatchNorm-ReLu. Further, their PatchGAN employs a single generator, whereas our approach employs two.  

\begin{figure}[!t]
\begin{center}
\includegraphics[width=\linewidth]{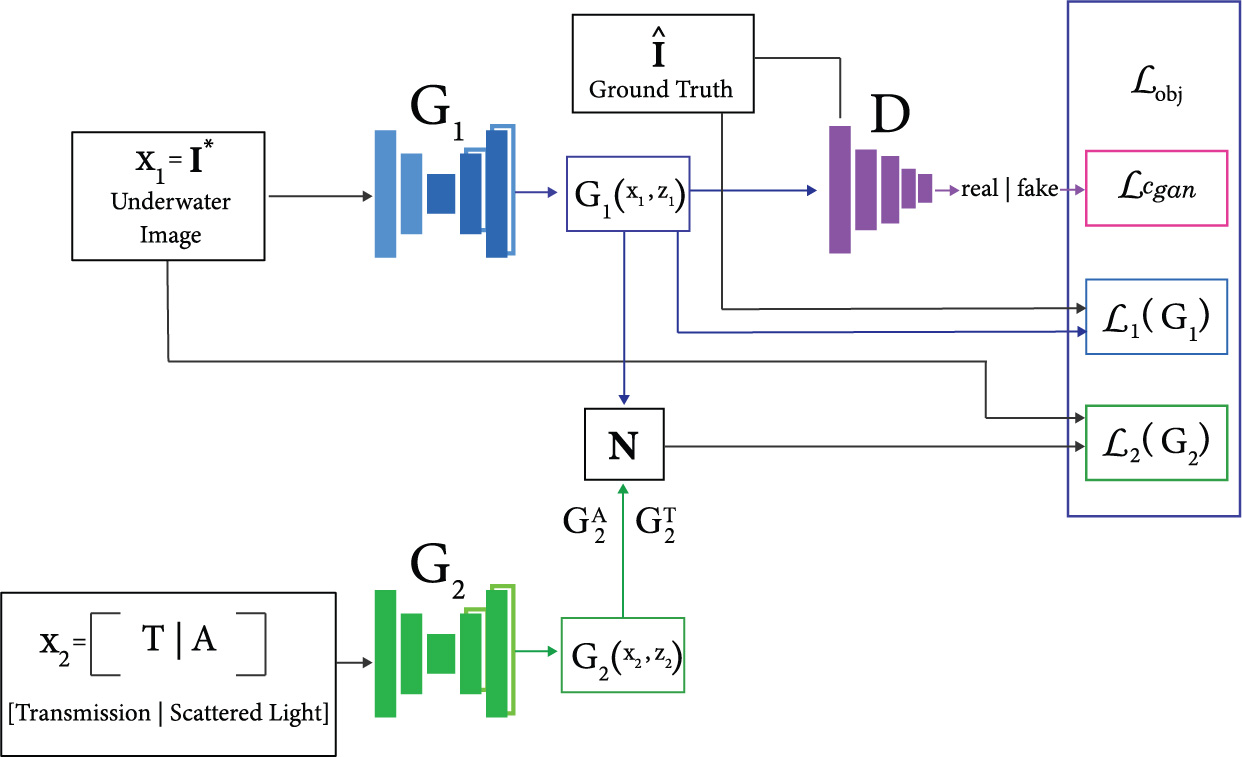} 
\end{center}
   \caption{Architecture of the proposed DGD-cGAN. It consists of two generators. The first $G_{1}$ dewaters the image. The second $G_{2}$ delivers $G_2^A$ and $G_2^T$ which are used to compute $\mathbf{N}$ following the UIFM in \cite{Duntley:1963}) making use of Equation \ref{eq6}. Note the computation of $\mathbf{N}$ involves the output of the first generator $G_1$, with the corresponding loss $\mathcal{L}_2$ for $G_2$ employing the input underwater image. The loss $\mathcal{L}_1$ for $G_1$ employs the ground truth image instead.
   }
\label{fig1}
\end{figure}

In Figure \ref{fig1}, we illustrate the architecture of DGD-cGAN for dewatering underwater images. In the figure, the first generator $G_{1}$ maps an underwater image input $x_{1}$ and a noise vector $z_{1}$ containing a hidden condition to the ground truth image $y_{1}$, i.e., $G_{1}:\{ x_{1},z_{1}\} \rightarrow y_{1}$. Image dewatering is hence performed by $G_1$ by predicting the output $G_1(x_{1},z_{1})$. The discriminator $D$ and the second generator $G_{2}$ both receive the output $G_1(x_{1},z_{1})$. Under this setting, the second generator $G_{2}$ will predict the transmission $G_2^{T}$ and the second term on the right-hand side of Equation \ref{eq3}, i.e. $G_2^{A}$, that contains the veiling light such that $G_{2}:\{ x_{2},z_{2}\} \rightarrow y_2$, where $y_2=[G_2^{T}\mid G_2^{A}]$.



This allows for the UIFM in Equation \ref{eq3} to recover a predicted underwater image from both $y_2$, and $y_1$. This, in turn allows for the formulation of a loss that aims to optimise two objectives. The first is pertaining the consistency of the dewatered image with the underwater scene as constrained by the UIFM in \cite{Duntley:1963}; the second is the colour restoration. The network objective is hence given by 
\begin{equation} \label{eq3.1}
\begin{split}
\min_{G_1} \max_D\mathcal{L}_{^cgan}(G_1,D)=&\mathbb{E}_{x_1,y_1}\left[ \log D(x_1,y_1)  \right]+\\&\mathbb{E}_{x_1,z_1}\left[ \log (1 - D(x_1, G_1(x_1,z_1))  \right],
\end{split}
\end{equation}
where the discriminator $D$ aims to maximize the function while $G_1$  tries to minimize it. 


For both generators, we use the L1-norm, denoted as $||\cdot||_1$. For our first generator $G_1$, the loss is given by
\begin{equation} \label{eq4}
\mathcal{L}_{1}(G_{1})= \mathbb{E}_{x_{1},\hat{\mathbf{I}},z_{1}}\left[||\hat{\mathbf{I}} - G_1(x_{1},z_{1})||_1 \right],
\end{equation}
where $x_1$ is the underwater image input and the output $G_1(x_{1},z_{1})$ corresponds to radiance image whose ground truth is $\hat{\mathbf{I}}$. As mentioned earlier, $z_{1}$ is the noise vector with the hidden condition, and it is related to $\hat{\mathbf{I}}$. 

Similarly, the loss for the second generator $G_2$ is given by
\begin{equation} \label{eq5}
\mathcal{L}_{2}(G_{2})= \mathbb{E}_{x_{2},\mathbf{I}^*,z_{2}}\left[||\mathbf{I}^* - \mathbf{N}||_{1}  \right],
\end{equation}
where $x_2=[\mathbf{T}\mid \mathbf{A}]$ is the matrix-wise concatenation of the light transmission and the veiling light inputs, and $z_2$ is the corresponding noise vector. In this loss, $\mathbf{I}^*$ = $x_1$ is the underwater image input and $\mathbf{N}$ is computed using the UIFM in Equation \ref{eq3}, so that
\begin{equation}\label{eq6}
\mathbf{N} = G_1(x_{1},z_{1})\odot G_2^{T} + G_2^{A}
\end{equation}
where the irradiance from the underwater scene is replaced by the output of $G_1$ and the transmission and veiling light taken from the unconcatenated outputs $G_2^{T}$ and $G_2^{A}$ of $G_2$. Note this becomes evident by using the short hands $\exp[-\alpha(z)r]$ for $G_2^{T}$,
$N(z_{t},\theta,\phi)\exp[+K(z,\theta,\phi)r \cos{(\theta)}]\times\{1-\exp[-\alpha(z)r+K(z,\theta,\phi)r \cos(\theta)]\}$ for $G_2^{A}$ and back-substituting them into Equation \ref{eq3}. 

Making use of the losses in Equations \ref{eq5} and \ref{eq6}, we can write the global loss as follows
\begin{equation} \label{eq9}
\mathcal{L}_{obj} = \min_{G_1} \max_D\mathcal{L}_{^cgan}(G_1,D) + \lambda_1\mathcal{L}_{1}(G_1)+\lambda_2\mathcal{L}_{2}(G_2),
\end{equation}
where $\lambda_1$ and $\lambda_2$ are constants that moderate the contribution of both generator objectives to the global loss.

\section{Experimental Setup}

\subsection{Implementation and Training}\label{implementation}

In this section, we ellaborate further upon our implementation and the datasets used for the results shown later on, in Section \ref{results}. Our implementation is based on PyTorch and all experiments shown here thereafter are run on a desktop with an NVIDIA GeForce RTX 2080 Ti\footnote{\label{note1}The code for our approach and its training is available at \url{https://github.com/SalPGS/DGD-cGAN}}. Note we use the Adam (Adaptive moment estimation) optimiser \cite{Kingma:2014} with a learning rate of $0.0002$ for $G_{1}$, $G_{2}$ and $D$. The momentum parameters are $\beta_1=0.5$ and $\beta_2=0.999$. To stabilize the training, we follow \cite{Masters:2018} and set the mini-batch gradient descent to 5 over 850 epochs. The loss weights are set to $\lambda 1 = 100.0$ and $\lambda 2 = 0.5$ for $\mathcal{L}_{1}(G_{1})$ and $\mathcal{L}_{2}(G_{2})$, respectively.

\subsection{Datasets}
\label{sct:datasets}
For training, we employ the TURBID dataset \cite{Duarte:2016} so as to allow for ground truth to validate the image dewatering and restoration. Due to the complexity of underwater environments, existing widely available datasets often do not include in-air ground truth. The TURBID dataset \cite{Duarte:2016} is a notable exception to this. It comprises ground truth, whereby the authors use a water tank to build a controlled underwater environment with a homogeneous illumination so as to apply incremental concentrations of milk, chlorophyll and a deep blue dye placing on the bottom human-made objects and seafloor elements such as rocks, stones and corals. Thus, the dataset contains RGB images of a turbid underwater scene and their clear water pair (i.e. ground truth). The RGB are divided into three categories: milk, deep blue and chlorophyll, with 20, 20 and 42 images, respectively. 

Note that  TURBID \cite{Duarte:2016} has a total of 81 high resolution underwater images with ground truth of size $3565\times2674$. These are a relatively small number, particularly when applied to neural network architectures, which often rely on a large number of examples for learning. To obtain the training data, we apply data augmentation by using different transformations on the data to increase its size and quality \cite{Khosla:2020}. To this end, we split all the images into four equal subimages of size $1807\times1355$ so as to later resize them to $256\times256$. This yields a total of 328 of turbid and ground truth image pairs. Once they are in hand, we apply a random vertical and horizontal flip  to diversify the dataset by changing the orientation of the images with probabilities of $0.5$ and $0.3$, respectively.  In all our experiments, the dataset is split into 80\% for training (295 images) and 20\% for testing (65 images) with size $256\times256$ for the first generator $G_{1}$. These images are trichromatic ones  with three colour channels. The second generator $G_{2}$ has two trichormatic inputs, the veiling light and transmission both of size $256\times256$ concatenated so as to comprise a six-dimensional input tensor. We ellaborate more on these in Section \ref{Light_transmission}


For testing, in addition to the TURBID dataset \cite{Duarte:2016}, we also use the Underwater Image Enhancement Benchmark (UIEB)  dataset \cite{Li:2019:WaterNet}, the Enhancing Underwater Visual Perception (EUVP) dataset \cite{Islam:2020}, the subset of ImageNet \cite{Deng:2009} (ImgN.-UG), the testing images from Sea-thru \cite{Akkaynak:2019} and the UFO-120 \cite{Islam:2020:UFO} dataset. Recall that UIEB dataset \cite{Li:2019:WaterNet} contains 950 underwater images, where 850 of them have their synthetic reference image enhanced by implementing 12 different techniques such as Retinex \cite{Fu:2014}, UDCP \cite{Drews:2016}, Red Channel \cite{Galdran:2015} and MSCNN \cite{Ren:2016} amongst others. The authors then selected the best quality enhanced image from the 12 techniques for each scene. The EUVP \cite{Islam:2020} dataset contains 12K paired underwater images of poor and good quality, and 8K unpaired images. For generating the pairs, the method in \cite{Fabbri:2018} was implemented for image enhancement. Then, good quality images were selected according to their colour, contrast, sharpness, and identification of the objects in the scene. The Sea-thru dataset  \cite{Akkaynak:2019} contains real underwater images and their depth map information whereas the UFO-120  dataset \cite{Islam:2020:UFO} contains low resolution distorted underwater images and its high resolution underwater pair. This comprises 1500 paired samples for training and 120 for benchmark evaluation.

\subsection{Light Transmission and Veiling Light Estimation}\label{Light_transmission}

Taking advantage of the available ground truth, we estimate the veiling light $\mathbf{A}$ and transmission $\mathbf{T}$ from Eq. \ref{eq1} so as to provide the inputs to $G_2$. Note that $\mathbf{A}$ can be viewed as a constant over the entire scene since, for TURBID \cite{Duarte:2016}, the authors have used homogeneous illumination. This assumption is not overly restrictive since scattering in underwater environments often results in scenes that are lit by diffuse sunlight. Thus, we can approximate the value of the veiling light using the grey world (GW) algorithm \cite{Finlayson:1998,Buenaposada:2001}. Our choice is based on the work presented in \cite{Ancuti:2017}, where the authors show that GW achieves a high performance on underwater images. Recall that GW estimates the radiance in a scene through colour normalisation, i.e. averaging the pixel intensity for each of the three RGB channels over the input image $\mathbf{I}$. Hence, the veiling light is identical for each of the pixels in the image, being given by the vector 
\begin{equation}\label{eq7}
\mathbf{a} = \bigg[\frac{\alpha}{n} \sum_{v\in \mathbf{I}} R(v), \frac{\tau}{n} \sum_{v\in \mathbf{I}} G(v), \frac{\gamma}{n} \sum_{v\in \mathbf{I}} B(v)\bigg],
\end{equation}
where $\alpha$, $\tau$, and $\gamma$ are channel dependent scalars and $n$ is the total number of pixels $v$ in the image. 

By casting $\mathbf{a}$ into a tensor of an appropriate size, we can obtain the transmission by solving $\mathbf{T}$ from Equation \ref{eq1}, which yields the expression
\begin{equation}\label{eq8}
\mathbf{T} = (\mathbf{I}-\mathbf{A})\oslash(\hat{\mathbf{I}}-\mathbf{A})
\end{equation}
where, $\oslash$ denotes the Hadamard, i.e. elementwise, division and, as before, $\hat{\mathbf{I}}$ is the ground truth image and $\mathbf{I}$ is the input underwater image.

\section{Results}
\label{results}
To this end, we first describe the metrics used and then present the results yielded by DGD-cGAN and their comparison to those state-of-the-art methods are discussed.
    
\subsection{Evaluation Metrics}

To verify the effectiveness of our dewatering cGAN, we measure the similarity between the restored and the ground truth images. Four metrics are employed: Euclidean Distance (ED) \cite{Liwei:2005,LiJing:2009}, Peak-Signal-to-Noise Ratio (PSNR) \cite{Hore:2010,Sara:2019}, Structural Similarity Index Measure (SSIM) \cite{ZhouW:2004}, and Underwater Image Quality Measure (UIQM) \cite{Panetta:2016}. To compare the variance of colours between the reference image (i.e. the ground truth) and the generated image, we use ED to measure the distance between the pixels intensities of each RGB channel and the average across the image. We use PSNR to measure the ratio between the maximum possible power of the image under consideration and the power of the noise affecting the quality of the image delivered by the methods under study. 

To find if there is any distortion or dynamic range compression in the dewatered image, we choose two metrics: SSIM and UIQM. Recall that SSIM aims at measuring the structural similarity between two images making use of known properties of the human visual system. SSIM compares normalised local image patch-pairs according to correlation, luminance and contrast. UIQM is also inspired upon the human visual system but does not require a reference. Rather it uses the colourfulness, sharpness and contrast of an underwater image.

\begin{figure}
	\begin{center}
	\begin{subfigure}{0.11\textwidth}
    \includegraphics[width=\textwidth]{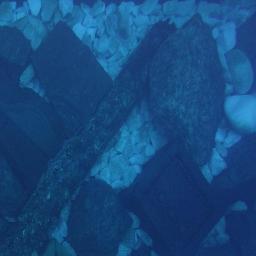}
    \includegraphics[width=\textwidth]{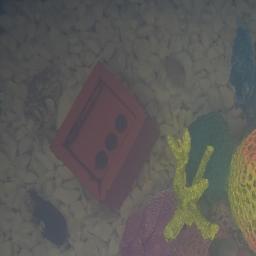}
    \caption*{\hspace{-1cm}\begin{tabular}{l}
    \multirow{2}{0.11\textwidth}{Input image}
    \end{tabular}}
\end{subfigure}
\begin{subfigure}{0.11\textwidth}
    \includegraphics[width=\textwidth]{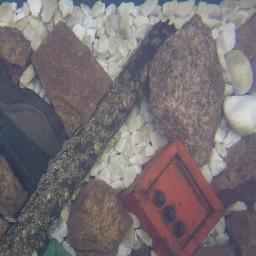}
    \includegraphics[width=\textwidth]{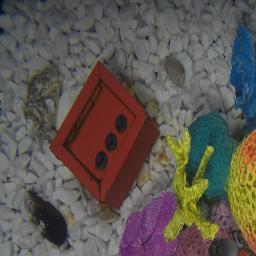}
    \caption*{\hspace{-1.3cm}\begin{tabular}{l}
    \multirow{2}{0.11\textwidth}{Ground truth}
    \end{tabular}}
\end{subfigure}
\begin{subfigure}{0.11\textwidth}
    \includegraphics[width=\textwidth]{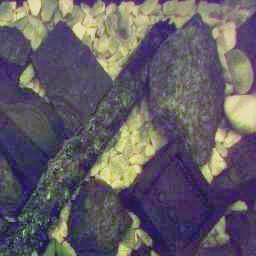}
    \includegraphics[width=\textwidth]{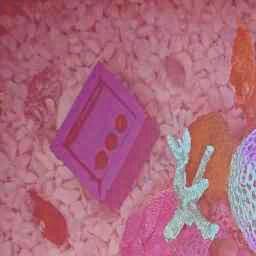}
    \caption*{\hspace{-1cm}\begin{tabular}{l}
    \multirow{2}{0.11\textwidth}{Retinex}
    \end{tabular}}
\end{subfigure}
\begin{subfigure}{0.11\textwidth}
    \includegraphics[width=\textwidth]{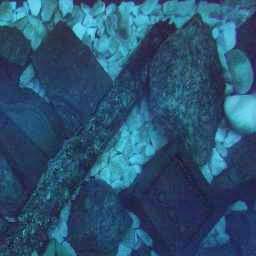}
    \includegraphics[width=\textwidth]{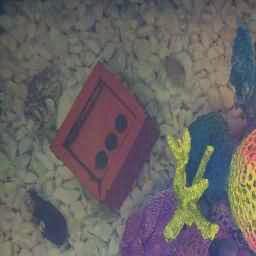}
    \caption*{\hspace{-1cm}\begin{tabular}{l}
    \multirow{2}{0.11\textwidth}{CLAHE}
    \end{tabular}}
\end{subfigure}
\begin{subfigure}{0.11\textwidth}
    \includegraphics[width=\textwidth]{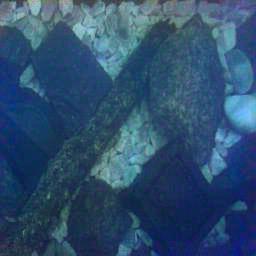}
    \includegraphics[width=\textwidth]{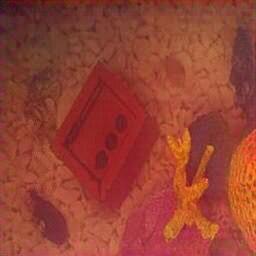}
    \caption*{\hspace{-1cm}\begin{tabular}{l}
    \multirow{2}{0.11\textwidth}{FUnIE-GAN}
    \end{tabular}}
\end{subfigure}
\begin{subfigure}{0.11\textwidth}
    \includegraphics[width=\textwidth]{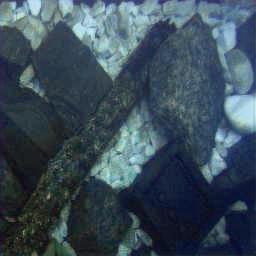}
    \includegraphics[width=\textwidth]{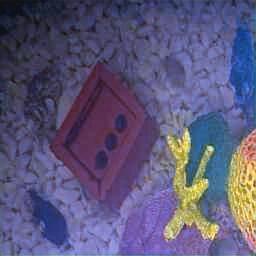}
    \caption*{\hspace{-1cm}\begin{tabular}{l}
    \multirow{2}{0.11\textwidth}{Water-GAN}
    \end{tabular}}
\end{subfigure}
\begin{subfigure}{0.11\textwidth}
    \includegraphics[width=\textwidth]{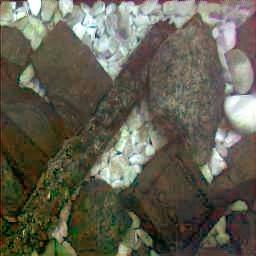}
    \includegraphics[width=\textwidth]{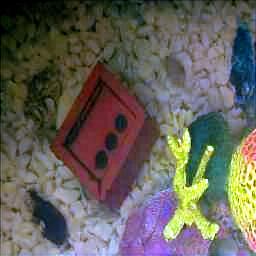}
    \caption*{\hspace{-0.9cm}\begin{tabular}{l}
    \multirow{2}{0.11\textwidth}{UGAN}
    \end{tabular}}
\end{subfigure}
\begin{subfigure}{0.11\textwidth}
    \includegraphics[width=\textwidth]{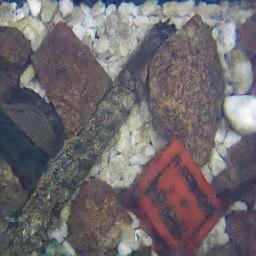}
    \includegraphics[width=\textwidth]{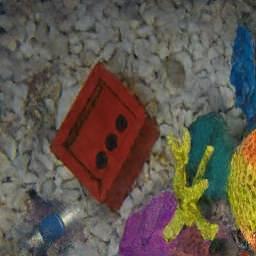}
    \caption*{\hspace{-0.8cm}\begin{tabular}{l}
    \multirow{2}{0.11\textwidth}{Ours}
    \end{tabular}}
\end{subfigure}
\end{center}
	\caption{Sample results on TURBID dataset imagery. From left-to-right: sample underwater input image from the dataset, the corresponding ground truth, and the results yielded by Retinex \cite{Jobson:1997}, CLAHE \cite{Zuiderveld:1994}, FUnIE-GAN \cite{Islam:2020}, Water-Net\cite{Li:2019:WaterNet}, UGAN \cite{Fabbri:2018} and DGD-cGAN.}
	\label{fig3}
\end{figure}

\newcolumntype{C}{>{\centering\arraybackslash}m{4em}}
\setlength{\tabcolsep}{1pt}
\renewcommand{\arraystretch}{1}
\begin{figure}[!t]
	\begin{center}
\begin{tabular}{cccccccc}
 {\rotatebox[origin=l]{90}{\scriptsize EUVP}} & \includegraphics[width=4.5em]{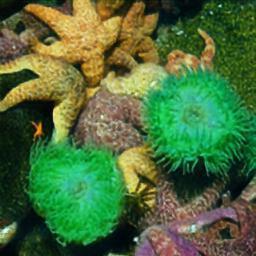}  & \includegraphics[width=4.5em]{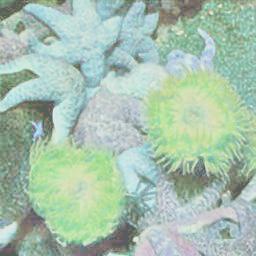}&\includegraphics[width=4.5em]{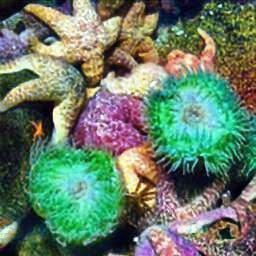} & \includegraphics[width=4.5em]{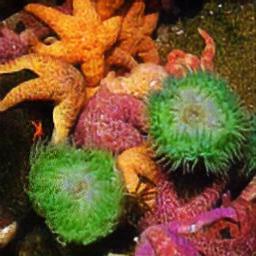}& \includegraphics[width=4.5em]{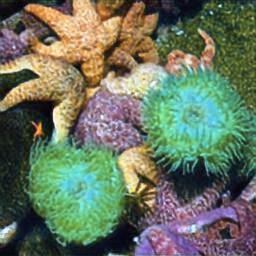}&\includegraphics[width=4.5em]{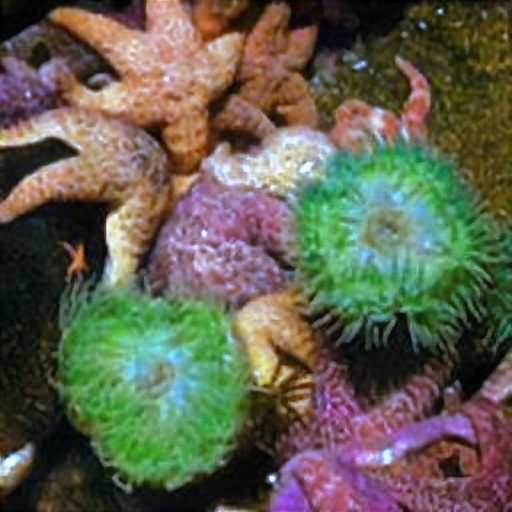}&\includegraphics[width=4.5em]{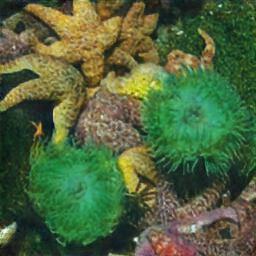}\\ 
{\rotatebox[origin=l]{90}{\scriptsize UIEB}} & \includegraphics[width=4.5em]{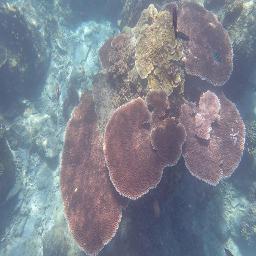} &\includegraphics[width=4.5em]{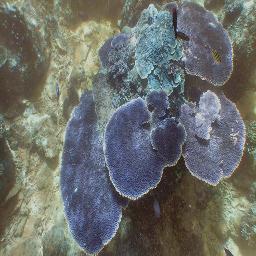} & \includegraphics[width=4.5em]{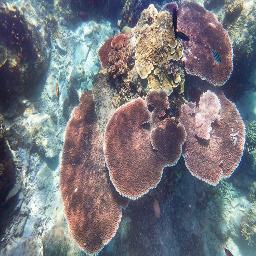}& \includegraphics[width=4.5em]{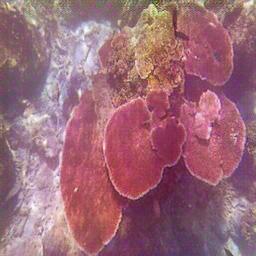} & \includegraphics[width=4.5em]{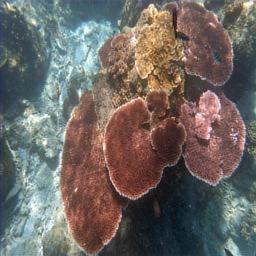} & \includegraphics[width=4.5em]{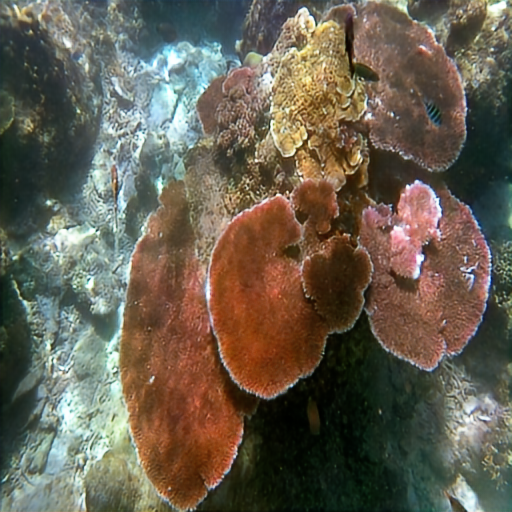} & \includegraphics[width=4.5em]{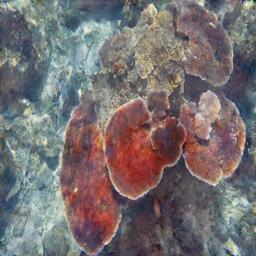}\\ 
{\rotatebox[origin=l]{90}{\scriptsize Sea-thru}} &\includegraphics[width=4.5em]{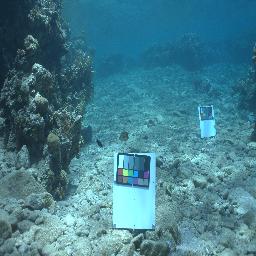} &\includegraphics[width=4.5em]{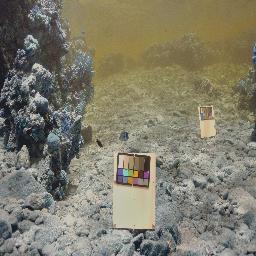} & \includegraphics[width=4.5em]{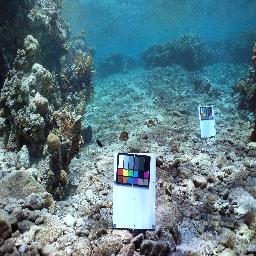}& \includegraphics[width=4.5em]{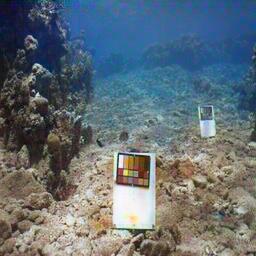} & \includegraphics[width=4.5em]{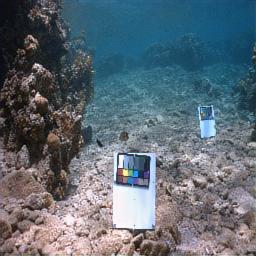} & \includegraphics[width=4.5em]{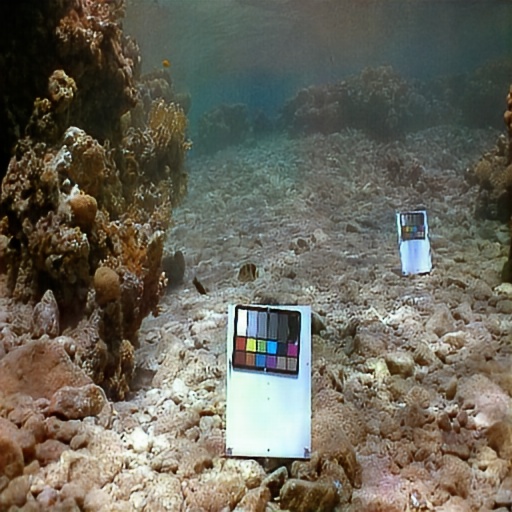} & \includegraphics[width=4.5em]{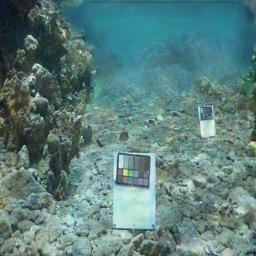} \\  
{\rotatebox[origin=l]{90}{\scriptsize ImageNet}} & \includegraphics[width=4.5em]{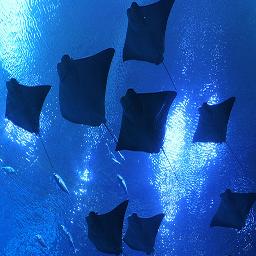} &\includegraphics[width=4.5em]{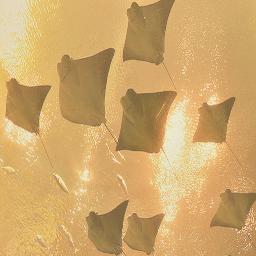} & \includegraphics[width=4.5em]{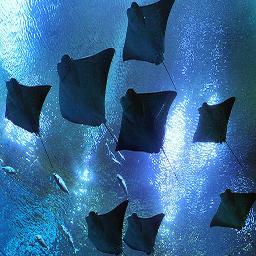}& \includegraphics[width=4.5em]{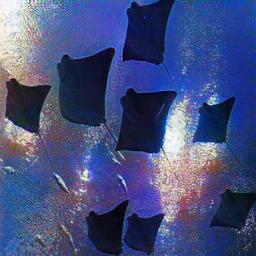} & \includegraphics[width=4.5em]{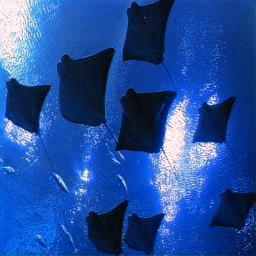} & \includegraphics[width=4.5em]{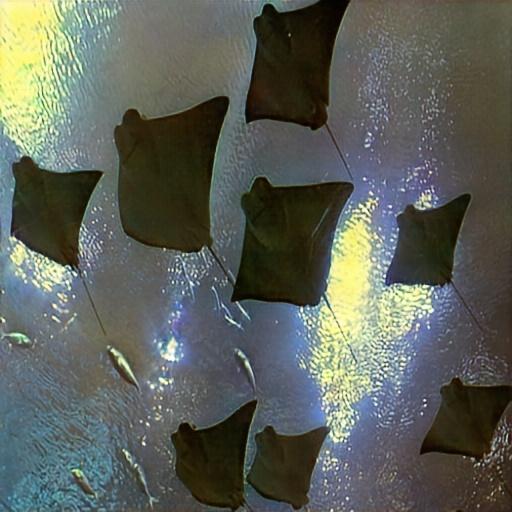} & \includegraphics[width=4.5em]{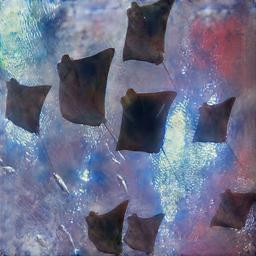} \\ 
& \scriptsize Input img. & \scriptsize Retinex & \scriptsize CLAHE & \scriptsize FUnIE& \scriptsize Water-Net & \scriptsize UGAN & \scriptsize Ours \\ 
\end{tabular}
	\end{center}
	\caption{Results on a sample image from datasets EUVP \cite{Islam:2020}, UIEB\cite{Li:2019:WaterNet}, Sea-thru \cite{Akkaynak:2019} testing images and ImageNet subset \cite{Deng:2009}. Each row, from left-to-right: the Underwater sample input image and the results yielded by Retinex \cite{Jobson:1997}, CLAHE \cite{Zuiderveld:1994}, FUnIE-GAN \cite{Islam:2020}, Water-Net\cite{Li:2019:WaterNet}, UGAN \cite{Fabbri:2018} and DGD-cGAN.}
	\label{fig4}
\end{figure}

\subsection{Comparison with SOTA Methods}
We now turn our attention to the effectiveness of DGD-cGAN for dewatering as compared with several state of the art (SOTA) methods, including two approaches for colour restoration, i.e. Retinex \cite{Jobson:1997} and CLAHE \cite{Zuiderveld:1994} and three deep learning methods for underwater image restoration and enhancement, i.e. Water-Net \cite{Li:2019:WaterNet}, FUnIE-GAN \cite{Islam:2020} and UGAN\cite{Fabbri:2018}.

Water-Net \cite{Li:2019:WaterNet} is a CNN-based approach for image enhancement. It implements HE \cite{Zuiderveld:1994}, WB \cite{Ancuti:2012} and GC to generate three inputs for each underwater image. A U-Net residual network receives these inputs together with the corresponding reference so as to predict a confidence map, which is then used to evaluate and select the features for the image enhancement operation. In our experiments, we use the pre-trained network provided by the authors\footnote{The code is available at \url{https://li-chongyi.github.io/proj_benchmark.html}}, which has been trained using the 800 paired images from UIEB dataset. FUnIE-GAN \cite{Islam:2020} is a cGAN trained on 11K paired images from EUVP dataset. It also comprises a modified U-Net and PatchGAN as the approach in \cite{Isola:2017}. The FUnIE-GAN \cite{Islam:2020} used here is a pre-trained one proposed by the authors\footnote{The code is available at \url{https://github.com/xahidbuffon/FUnIE-GAN}}, trained on 7.5K unpaired images from the same dataset and enhanced using CycleGAN \cite{Zhu:2017}. The third method, UGAN \cite{Fabbri:2018} uses a subset from the well-known ImageNet dataset \cite{Deng:2009} for training by generating paired images using a CycleGAN \cite{Zhu:2017}. The enhancement network is reminiscent of that in \cite{Isola:2017}, using a U-Net generator and a PatchGAN discriminator. The main difference is the adapted gradient penalty used in its training, which is similar to that in \cite{arjovsky2017wasserstein}. The network used here is, again, a pre-trianed version that has been made widely available by the authors\footnote{The code is available at \url{https://github.com/IRVLab/UGAN}}.

\begin{figure}[!t]
	\begin{center}
\begin{subfigure}{0.11\textwidth}
    \includegraphics[width=\textwidth]{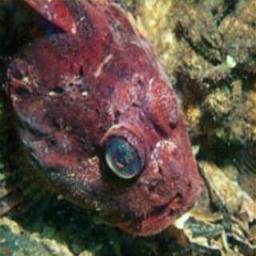}
   \includegraphics[width=\textwidth]{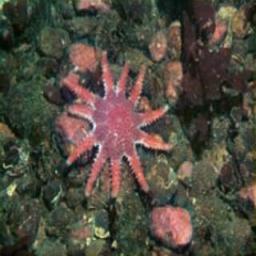}
    \caption*{\hspace{-1cm}\begin{tabular}{l}
    \multirow{2}{0.11\textwidth}{Input image}
    \end{tabular}}
\end{subfigure}
\begin{subfigure}{0.11\textwidth}
    \includegraphics[width=\textwidth]{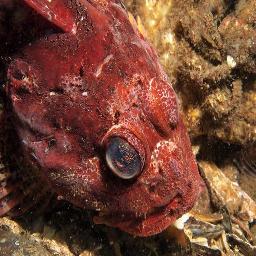}
    \includegraphics[width=\textwidth]{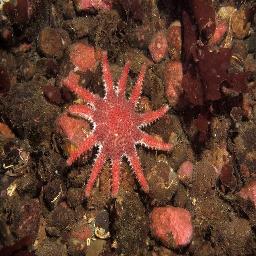}
    \caption*{\hspace{-1.3cm}\begin{tabular}{l}
    \multirow{2}{0.11\textwidth}{Benchmark}
    \end{tabular}}
\end{subfigure}
\begin{subfigure}{0.11\textwidth}
    \includegraphics[width=\textwidth]{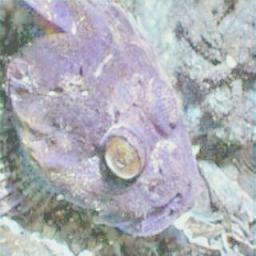}
    \includegraphics[width=\textwidth]{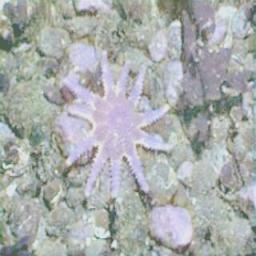}
    \caption*{\hspace{-1cm}\begin{tabular}{l}
    \multirow{2}{0.11\textwidth}{Retinex}
    \end{tabular}}
\end{subfigure}
\begin{subfigure}{0.11\textwidth}
    \includegraphics[width=\textwidth]{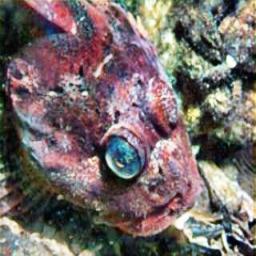}
    \includegraphics[width=\textwidth]{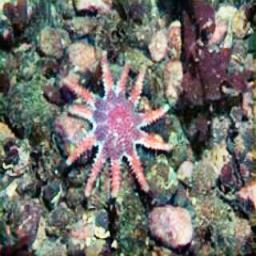}
    \caption*{\hspace{-1cm}\begin{tabular}{l}
    \multirow{2}{0.11\textwidth}{CLAHE}
    \end{tabular}}
\end{subfigure}
\begin{subfigure}{0.11\textwidth}
    \includegraphics[width=\textwidth]{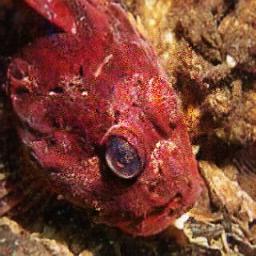}
    \includegraphics[width=\textwidth]{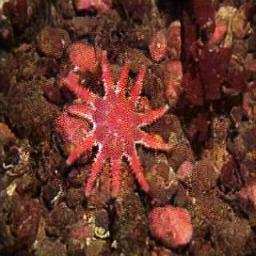}
    \caption*{\hspace{-1cm}\begin{tabular}{l}
    \multirow{2}{0.11\textwidth}{FUnIE-GAN}
    \end{tabular}}
\end{subfigure}
\begin{subfigure}{0.11\textwidth}
    \includegraphics[width=\textwidth]{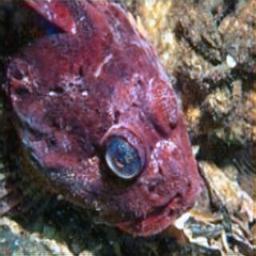}
    \includegraphics[width=\textwidth]{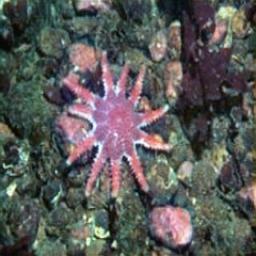}
    \caption*{\hspace{-1cm}\begin{tabular}{l}
    \multirow{2}{0.11\textwidth}{Water-GAN}
    \end{tabular}}
\end{subfigure}
\begin{subfigure}{0.11\textwidth}
    \includegraphics[width=\textwidth]{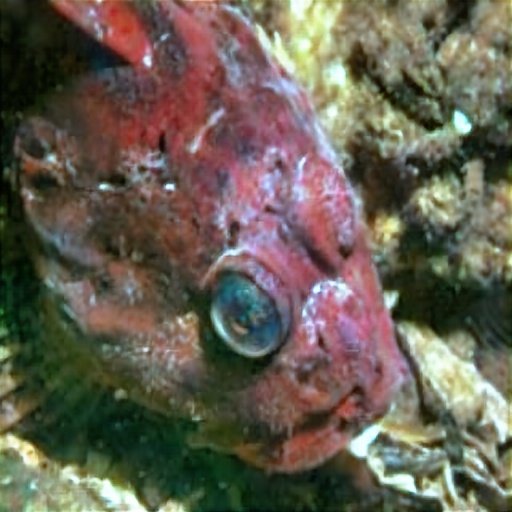}
    \includegraphics[width=\textwidth]{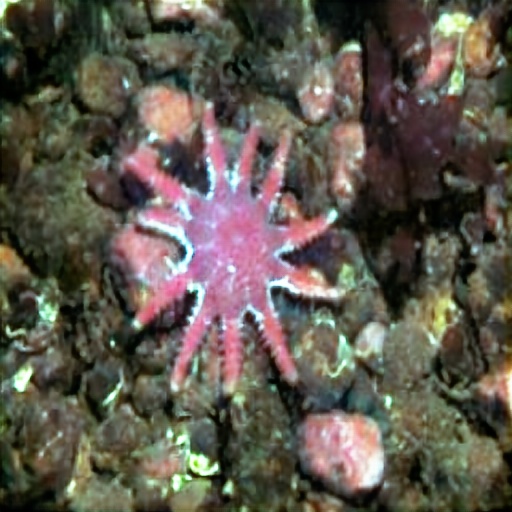}
    \caption*{\hspace{-0.9cm}\begin{tabular}{l}
    \multirow{2}{0.11\textwidth}{UGAN}
    \end{tabular}}
\end{subfigure}
\begin{subfigure}{0.11\textwidth}
    \includegraphics[width=\textwidth]{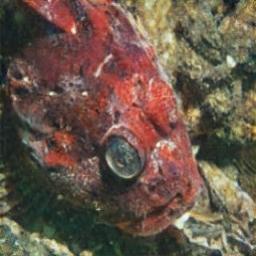}
    \includegraphics[width=\textwidth]{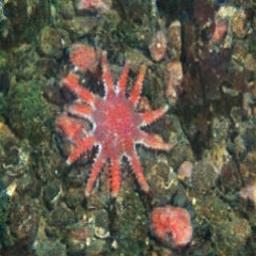}
    \caption*{\hspace{-0.8cm}\begin{tabular}{l}
    \multirow{2}{0.11\textwidth}{Ours}
    \end{tabular}}
\end{subfigure}
\end{center}
	\caption{Example results on the UFO-120 \cite{Islam:2020:UFO} dataset. From left-to-right: sample underwater input image from the dataset, high resolution image contained in the UFO-120 benchmark \cite{Islam:2020:UFO}, and results yielded by Retinex \cite{Jobson:1997}, CLAHE \cite{Zuiderveld:1994}, FUnIE-GAN \cite{Islam:2020}, Water-Net\cite{Li:2019:WaterNet}, UGAN \cite{Fabbri:2018} and DGD-cGAN.}
	\label{fig5}
\end{figure}

\begin{table}[!t]
  \begin{center}
    {\small{
\begin{tabular}{ccccc}
\toprule
Methods & $R\downarrow$   &  $G\downarrow$ & $B\downarrow$ & $RGB$ $Avg.\downarrow$\\
\midrule
        CLAHE & 0.1216& 0.1225&0.1170& 0.1203\\
		Retinex &  0.1700 & 0.1332& 0.1936& 0.1656\\
		Water-Net & 0.1233 &0.1136& 0.1160&0.1176\\
		FUnIE-GAN & 0.1500 & 0.1090 & 0.1395 & 0.1328\\
		UGAN& 0.1428 &  0.1502 & 0.1719 & 0.1549 \\
		DGD-cGAN &  \textbf{0.0814}& \textbf{0.0856} &\textbf{ 0.0902} & \textbf{0.0857}\\ 
\bottomrule
\end{tabular}
}}
\end{center}
\caption{Euclidean distance (ED) \cite{Liwei:2005,LiJing:2009} for TURBID dataset \cite{Duarte:2016} yielded by Retinex \cite{Jobson:1997}, CLAHE \cite{Zuiderveld:1994}, FUnIE-GAN \cite{Islam:2020}, Water-Net \cite{Li:2019:WaterNet}, UGAN \cite{Fabbri:2018} and DGD-cGAN. The table lists both, the ED for each of the RGB channels and the average across them. The absolute best results are in bold. The lower the distance, the better the result.}
\label{tab: Euclidean distance}
\end{table}

\begin{table}[!t]
  \begin{center}
    {\small{
\begin{tabular}{ccc}
\toprule
Methods & $SSIM\uparrow$ & $PSNR\uparrow$  \\
\midrule
		Retinex & 0.6557 & 17.1998 \\
		Clahe & 0.6983 & 20.4727\\
		UGAN & 0.6216 & 18.0814\\
		Water-Net & 0.6792 & 19.5851\\
		FUnIE-GAN & 0.6477 &  22.3152 \\
		DGD-cGAN & \textbf{0.7061} & \textbf{23.8426}\\
\bottomrule
\end{tabular}
}}
\end{center}
\caption{SSIM and PSNR metrics on TURBID dataset \cite{Duarte:2016} between the imagery yielded by each of the methods under consideration and the ground truth in the dataset. The absolute best results are in bold. The higher the SSIM and PSNR, the better the result.}
\label{tab: SSIM and PSNR}
\end{table}

In our experiments, we note that the UQIM metric \cite{Panetta:2016} does not require a reference image since its evaluation is done between the underwater image and the restored scene on a perceptual basis. Thus, UQIM can be used for all the datasets under consideration, regardless of whether they provide in-air ground truth or not. For PSNR, SSIM and ED, the evaluation is performed on the TURBID dataset as described in Section \ref{sct:datasets}.

We commence by showing qualitative results of all the methods under consideration in Figures \ref{fig3} - \ref{fig5}. Sample imagery for TURBID dataset and UFO-120 dataset are shown in Figure \ref{fig3} and Figure \ref{fig5}, respectively. Figure \ref{fig4} shows sample results for the other datasets under consideration.  Note that our method produces imagery that is in good accordance with the ground truth in Figure \ref{fig3}, preserving the detail missed by some of the alternatives and without introducing unwanted orange or blue tinges. 

\begin{table}[!t]
  \begin{center}
    {\small{
\begin{tabular}{ccccccc}
\toprule
Methods & EUVP& UIEB & TURBID& Sea-thru& ImgN.-UG.& UFO-120\\
\midrule
Baseline & 2.2936& 2.9073& 2.0296&2.4455&2.4185&2.5829\\
Clahe & 2.6693& 3.0767& 2.5487&2.5180&2.6263 &2.9449\\
FUnIE-GAN & 2.9629&  3.2420& 2.7726&2.6547&2.8677&2.8925\\
Retinex & 3.0067& 3.1325& 2.9552 & 3.2596&3.0924&3.0014\\
Water-Net & 3.0167& 3.2234&  2.8869& 3.0367&3.0584&3.0787\\
UGAN & 2.9993 & 3.0742 & 2.8959& 2.8429& 3.0862&3.0164\\
DGD-cGAN & \textbf{3.0829}& \textbf{3.4184}& \textbf{3.2718}&\textbf{3.1928}&\textbf{3.1756}&\textbf{3.1885}\\

\bottomrule
\end{tabular}
}}
\end{center}
\caption{UQIM metric values of the methods on datasets under consideration. As UQIM measures the quality of image based upon perceptual criteria devoid of the need for ground truth or reference imagery, we also show the UQIM of the input imagery for each dataset as a baseline. The higher the UQIM, the better the result.}
\label{tab: UQIM}
\end{table}

We also show quantitative results in Tables \ref{tab: Euclidean distance}, \ref{tab: SSIM and PSNR} and \ref{tab: UQIM}. Table \ref{tab: Euclidean distance} shows the Euclidean distance (ED) between the images yielded by the methods under consideration and the ground truth in the TURBID dataset \cite{Duarte:2016}. Since the ED effectively measures the colour similarity between the predicted and ground truth, the lower the Euclidean distance, the better the result. In this table, we show both the per-channel and the trichromatic ED, with the absolute best results in bold font. Note that our approach outperforms all the alternatives, both on a per-channel basis and on average across the image colour. 

In Table \ref{tab: SSIM and PSNR}, we show the corresponding results of PSNR and SSIM for the TURBID dataset \cite{Duarte:2016}, with the absolute best results denoted in bold. Again, our approach outperforms the alternatives, with a higher PSNR and SSIM, with CLAHE \cite{Zuiderveld:1994} being the second best for the SSIM and FUnIE-GAN  \cite{Islam:2020} for the SSIM.

Finally, in Table \ref{tab: UQIM} we show the UQIM \cite{Panetta:2016} per-method and dataset under study. Since the UQIM does not require a reference or ground truth image but rather assessing the quality of the imagery based upon perceptual metrics, we include the score for the input image per dataset as a baseline. As before, in Table \ref{tab: UQIM}, we denote the absolute best results in bold font. Our DGD-cGAN is the absolute best across all datasets, with Water-Net \cite{Li:2019:WaterNet} being the second best for the EUVP, UIEB and UFO-120 datasets. Somewhat surprisingly, Retinex is the second best for the Sea-thru and ImgN.-UG. datasets. These results are consistent with those shown in the other tables and our qualitative results shown in Figures \ref{fig3}, \ref{fig4} and \ref{fig5}.



\section{Conclusions}
In this paper, we introduce a conditional generative adversarial network
(cGAN) with two generators for image dewatering. Our Dual Generator Dewatering cGAN (DGD-cGAN) can remove the colour cast of underwater images and restore their true colours. The first generator predicts the dewatered image, while the second learns the transmission and veiling light components of the underwater image formation model. We also show how a loss function which minimises two objectives can be formulated with these two generators. The loss function presented aims at minimising the inconsistency of the dewatered image with their underwater scene as constrained by the UIFM in \cite{Duntley:1963} and the colour difference between the ground-truth and the in-air image. We evaluate our approach using several widely available datasets and compare the results with SOTA methods elsewhere in the literature. Our experiments show, both quantitatively and qualitatively, that DGD-cGAN consistently delivers a margin of improvement over the alternatives.

\bibliographystyle{unsrt}  
\bibliography{references}

\end{document}